# Superconducting and thermal properties of *ex-situ* GlidCop® sheathed multifilamentary MgB$_2$ wires

A Malagoli[1], M Tropeano[1,2], V Cubeda[2], C Bernini[1], V Braccini[1], C Fanciulli[1], G Romano[1,2], M Putti[1,2] and C Ferdeghini[1]

[1] CNR-INFM LAMIA, Corso Perrone 24, 16152 Genova Italy
[2] Physics Department, University of Genova, Via Dodecaneso 33, 16146 Genova Italy

**Abstract**
In DC and AC practical applications of MgB$_2$ superconducting wires an important role is represented by the material sheath which has to provide, among other things, a suitable electrical and thermal stabilization. A way to obtain a large enough amount of low resistivity material in to the conductor architecture is to use it as external sheath.
In this paper we study *ex-situ* multifilamentary MgB$_2$ wires using oxide-dispersion-strengthened copper (GlidCop®) as external sheath in order to reach a good compromise between critical current density and thermal properties. We prepared three GlidCop® samples differing by the content of dispersed sub-microscopic Al$_2$O$_3$ particles. We characterized the superconducting and thermal properties and we showed that the good thermal conductivity together the good mechanical properties and a reasonable critical current density make of GlidCop® composite wire a useful conductor for applications where high thermal conductivity is request at temperature above 30K, such as Superconducting-FCL.

**1.Introduction**
In the last years the study and development of MgB$_2$ have made of it a good material for several electric power applications. Indeed, the significant improvement of its superconducting properties has been accompanied by an industrial development of the technology to produce long length tapes and wires [1, 2].
Furthermore, if we consider that such superconductor can operate between 20K and 30K, within the range of a commercial cryocooler, and that its raw materials and production costs are much lower than those of the high T$_c$ superconductors, then it is easy to think at MgB$_2$ as a useful conductor for magnets, fault current limiters, motors, generators and transformers [3].
In all these practical applications a fundamental role is played by the material sheath and architecture of the conductor since they have to provide suitable mechanical properties for windings, protection against magnetic flux jumps and an appropriate thermal and electrical stabilization [4, 5, 6]. These issues lead to focus the conductor development on the multifilamentary wires or tapes including a material with low resistivity and high thermal conductivity.
Such MgB$_2$ conductors have been developed using both the *in-situ* and *ex-situ* PIT method; in particular regarding the *in-situ* technique the use of copper (Cu) and oxide-dispersion-strengthened copper (ODS Cu) as stabilizing material was investigated [7]. In some cases the Cu or the ODS Cu were used as inner core of a multifilamentary wire, in other cases as external sheath. Concerning the critical current density, the best results were obtained on the samples with the Cu or ODS Cu placed in the inner part.
Also for the *ex-situ* method the issue has been faced: Columbus Superconductors in collaboration with CNR-INFM LAMIA, has developed a stabilized multifilamentary tape with the inner part of oxygen free high conductivity Copper (OFHC Cu) [2].
Still such configuration is not the optimal solution because a desirable increasing of the copper cross section to improve the stabilization properties should go to the detriment of the number of filaments and vice versa. The total cross section being equal, it could be preferable to have the stabilizing material in the outer part of the conductor. As it is known, the *ex-situ* way is based on the using of hard enough sheaths, [8, 9] as Nickel or Nickel-alloys, but these materials have high resistivity and low thermal conductivity, therefore they are bad stabilizers. In these cases the only way to have the stabilizing material on the outside of the conductor is to make a further process called "wire-in-channel", where the superconducting wire is inserted and soldered into a copper channel. On the other side the copper being too soft metal, it is not an adequate material for the external sheath in *ex-situ* wires. Focusing our efforts on the *ex-situ* multifilamentary wires we have tried to use the ODS Cu as external sheath since it seems to be a good compromise between electrical/thermal and mechanical properties.
We have prepared several samples using three different kinds of the so called GlidCop®, i.e. with 0.15%, 0.25% and 0.60% Al$_2$O$_3$ weight. In this work we present the superconducting properties characterization of such wires as well as their thermal conductivity measurements.

## 2. Experimental details

Three GlidCop® samples, which differ by the content of $Al_2O_3$ nanoparticles, i.e. 0.15%, 0.25% and 0.60% in weight, and one OFHC Cu sample were prepared with the aim to compare their mechanical and electrical properties. In particular we have fabricated four tapes with a thickness of 0.25 mm and a width of 4 mm and they underwent the same heat treatment used in the *ex-situ* PIT method to prepare $MgB_2$ conductors. For these samples, resistivity and stress vs strain measurements were carried out with a commercial PPMS system by Quantum Design and by an INSTROM system, respectively.

In a second step, multifilamentary wires have been prepared by means of the *ex-situ* PIT method. A Nickel tube was filled by homemade $MgB_2$ powders. The tube was cold worked to obtain a monofilamentary wire with a diameter of about 2 mm. 19 pieces of this wire were inserted into a GlidCop tubes. Further cold working has been performed to obtain a square wire of about 1,76 x 1,76 $mm^2$. As external sheath, we used three GlidCop® tubes, one for each type. The three samples were called AL15, AL25 and AL60 respectively. The samples underwent a final heat treatment at about 950°C using a continuous heat treatment system.

The manufactured $MgB_2$ nultifilamentary wires have been characterized by means of inductive critical current density measurements at 5K and 20K in an applied magnetic field up to 5 Tesla using a commercial SQUID system developed by Quantum Design.

Thermal conductivity was evaluated with the Thermal Tansport Option (TTO) of a Physical Property Measurements System (PPMS) by Quantum Design.

We have compared the results with those obtained on a wire with the same architecture but with MONEL as external sheath [10] and with a standard tape produced by Columbus Superconductors SpA.

Finally, we want to point out that our purposse is to compare the material sheaths, therefore the $MgB_2$ powders used here were standard homemade powders [11], without any optimization by doping [2], ball milling [2] or particular heat treatment [10].

## 3. Results

*3.1. Mechanical and electrical characterization of GlidCop® specimens: a comparison with OFHC Copper.*

First of all we have characterized the three different GlidCop® materials and the OFHC Cu carrying out stress vs strain and resistivity vs temperature measurements on samples that have undergone similar cold working and the same heat treatment scheduled by the *ex situ* technique to fabricate $MgB_2$ conductors.

In figure 1 the stress vs strain measurements for the four samples are shown.

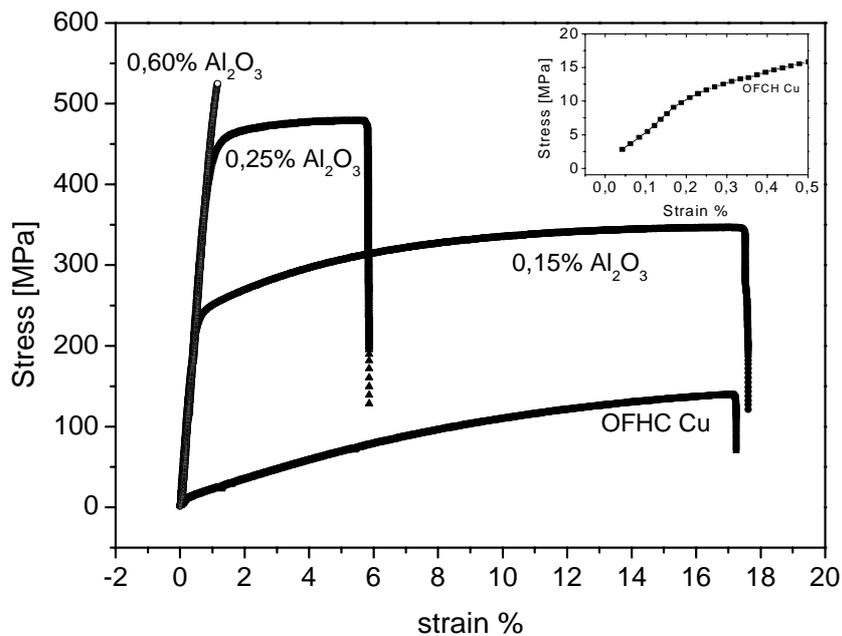

Figure1: stress vs strain measurements for the four samples. In the in-set the stress at which the OFHC Cu begins to deform plastically is highlighted.

As it can be seen increasing the amount of $Al_2O_3$ the toughness of the GlidCop® increases. For these samples the stress at which the material begins to deform plastically (yield point) reaches values more than one order of magnitude higher than OFHC Cu that is highlighted in the in-set of the figure 1. We would like to notice that for the sample AL60 the toughness is higher than 550 MPa, the maximum value that can be measured by our instrument. The resistivity data (RRR and $\rho_0$) of the four samples are reported in table I. As expected, with the increasing of the $Al_2O_3$ amount, the RRR decreases and the resistivity increases. It is remarkable the low value of RRR namely for the OFCH Cu sample. This can be explained considering that this material contains a dispersion of oxidable particles able to prevent the copper oxidation or an oxide diffusion. The heat treatment at high temperature could cause a reaction of these particles as well as the $Al_2O_3$ particles with the copper matrix decreasing the relative RRR. However using the GlidCop® it is possible to obtain a toughness improvement of two order of magnitude keeping a resistivity not too different from the OFCH Cu one, as in the case of the AL15 sample.

| Sample | $\rho_0$ ($\mu\Omega cm$) | RRR |
|---|---|---|
| OFHC Cu | 0.041 | 40 |
| AL15 | 0.106 | 15 |
| AL25 | 0.214 | 9 |
| AL60 | 0.315 | 7 |

Table I: resistivity data of the four samples

*3.2. Superconducting properties of multifilamentary samples*
In figure 2 the cross sections of the AL15, AL25 and AL60 $MgB_2$ multifilamentary samples are shown. As can be seen the different toughness is reflected on the capability to cold work the Nickel-sheathed monofilaments, indeed the AL60 cross section, concerning the filaments structure, is more regular and the filaments they self are more homogeneous.

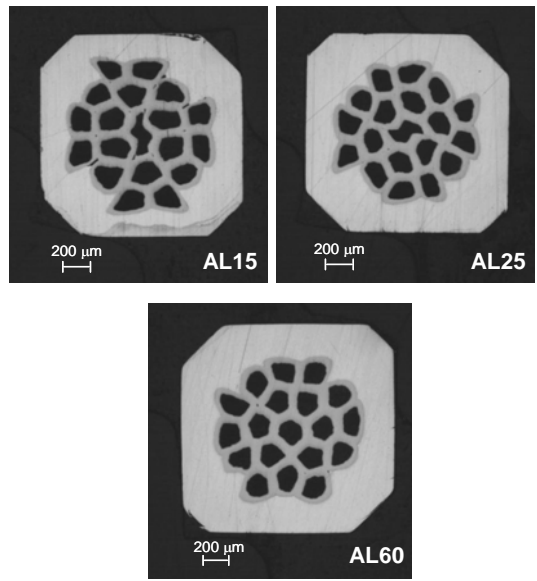

Figure 2: Cross sections of the three GlidCop® sheathed $MgB_2$ multifilamentary wires

In spite of the high temperature (between 900 and 1000 °C) of the final heat treatment, there is not an evident diffusion of the Copper through the Nickel, indeed any filament seems to be poisoned by Copper. Probably this is thank to the shortness of the heat treatment process performed by our continuous system. As a consequence, it is not needed to insert a further chemical barrier, as for example Niobium, to protect the $MgB_2$ filaments.
In figure 3 the magnetic $J_C$ measurements at 5K are reported.

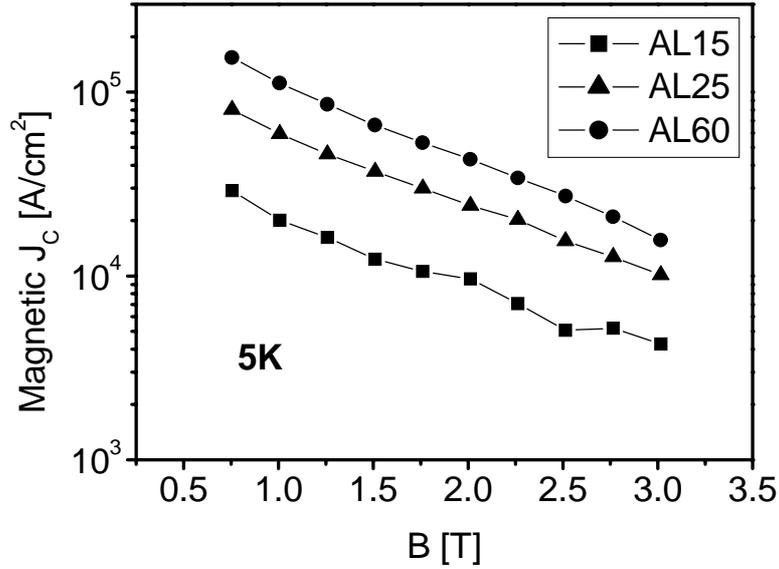

Figure 3: Magnetic $J_C$ behavior in applied magnetic field at 5K

Also concerning the $J_C$, its values increase with the increasing of the external sheath toughness. The higher values were obtained with the AL60 sample while the $J_C$ behavior in the magnetic field is the same for the three samples. This suggests that the effect of the higher external sheath toughness in the AL60 sample is to increase the $MgB_2$ powder packing inside the filaments and consequently to improve their connectivity without any change of the pinning mechanism..

From an applicative point of view it is important to test the superconducting wires at a temperature of 20 K because, as it is known, it can be easily obtained by now using commercial and not expansive cryocoolers.

In figure 4 the behavior of $J_C$ in magnetic field at 20 K for the AL60 sample is shown. For comparison, we have also reported the $J_C$ values of two other samples: a MONEL sample which is a multifilamentary wire with the same architecture and obtained with the same procedure than the AL60 one, except for the external sheath material [10], and a OFHC Copper stabilized multifilamentary tape produced by Columbus Superconductors. All these wires have the same $MgB_2$ powders. The MONEL sample has an harder external sheath while the standard tape has a stabilizing core.

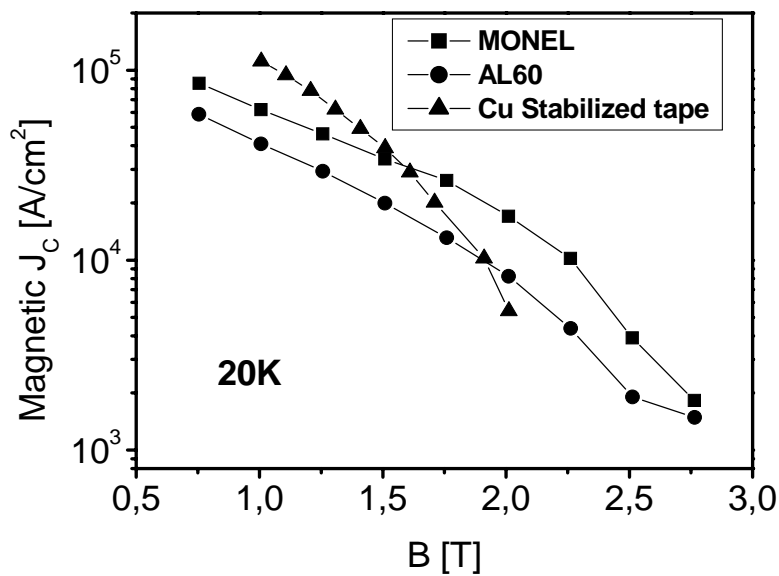

Figure 4: $J_C$ behavior in magnetic field at 20 K: comparison between AL60, MONEL and Cu Stabilized tape samples

Both wires have at low field lower $J_C$ values than the tape; this is due to the fact that the tape underwent a further flat rolling deformation and therefore the $MgB_2$ is more compacted. Still the AL60 and MONEL wires have a better $J_C$ behavior at magnetic field higher than 2 Tesla. Being the tape in perpendicular configuration, at these magnetic fields, an important role is played by the anisotropy, which is present in the tape but not in the wires.

*C. Thermal conductivity*

In figure 5 the thermal conductivity K of the specimens are shown in the temperature range 5-200K. Here are reported K the three GlidCop® multifilamentary samples, K of OFHC copper stabilized Columbus tape and K of the MONEL wire. All are normalized to the whole sample cross sections. In such a way the K values clearly depend on the amount of the metal elements in the wire.

The thermal conductivity behavior of the stabilized tape is similar to the typical trend of pure metals [12], with a maximum value of 220 (W/m K) at 20K, and this is clearly related to the presence of the OFHC Copper for the stabilization of the tape. For the three GlidCop® wires, the thermal conductivity reach different peak values at higher different temperatures with increasing $Al_2O_3$ content in the matrix, before all converge on the same curve above 100K.

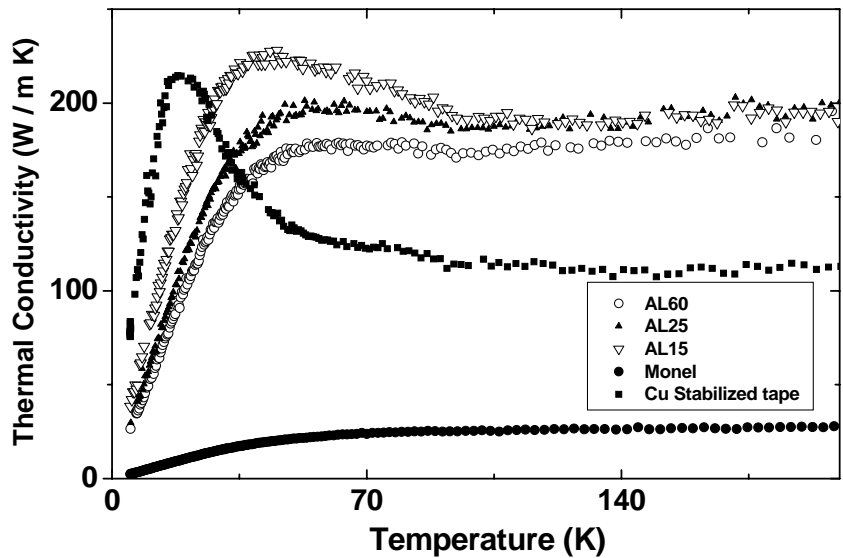

Figure 5: Thermal conductivity of the three GlidCop® multifilamentary samples compared with a OFHC Copper stabilized Columbus multifilamentary tape and the Monel wire .

For the AL15 sample the thermal conductivity reaches a maximum value of 230 W/m K at about 40K. Increasing the amount of $Al_2O_3$ nanoparticles in the AL25 and AL60 samples results in a reduction of the maximum value of the thermal conductivity (200W/m K and 190 W/m K respectively) and a shift of the peak toward higher temperatures (55K and 65K respectively).

The thermal conductivity of the reference MONEL wire is quite low, and it doesn't present, as expected for an alloy, any low temperature maximum.

For all the samples the thermal conductivity is almost constant in the range 130-300K. Assuming that thermal conductivity complies with rule of mixture, we can calculate the thermal conductivity of the samples as

$$K_{tape} = \Sigma\, f_i\, K_i \qquad (1)$$

where $f_i = S_i/S$, with $S_i$ the surface of the i sheath and S the total surface, and $K_i$ are the respective thermal conductivity.

| Wire or tape | MgB$_2$ f$_i$ % | GlidCop f$_i$ % | Monel f$_i$ % | Nickel f$_i$ % | Iron f$_i$ % | Copper f$_i$ % | K$_{300}$ (W/K m) calculated | K$_{300}$ (W/K m) measured |
|---|---|---|---|---|---|---|---|---|
| Copper stabilized | 10 | - | - | 67 | 8 | 15 | 128 | 112 |
| GlidCop | 16-18 | 58 | - | 26-24 | - | - | 212 | 200 |
| Monel_Ni | 30 | - | 45 | 25 | - | - | 36 | 27 |

Table2: f$_i$ values for the elements in the metal matrix of wire or tape, K calculated and K measured at 300K for the wire and tape examined. The K$_{300K}$ is evaluated by eq.(1) with K$_{Nickel}$=91W/m K, K$_{Copper}$ = 400W/m K, K$_{Monel}$ = 22W/m K and K$_{Iron=}$ 80W/mK (ref [14]); the K$_{GlidCop}$= 322 is from North American Hoganas Data Sheet; the value for K$_{MgB2}$=10 is from ref [15].

In table 2 the f$_i$ value of the different sheath components for the various wires are reported, together with the measured and calculated thermal conductivity according to (1). We can notice that the calculated and measured K$_{300K}$ values agree rather well even if the calculated are slightly higher than the measured ones; the differences can be related to the fact that we don't take in account the reaction layer between the MgB$_2$ filaments and the Nickel sheath [13]. This layer presents high resistivity and therefore low thermal conductivity so reducing the cross sectional area of the higher conducting element.

## 4. Conclusions

In this work the possibility of using GlidCop® as external sheath in *ex-situ* MgB$_2$ wires has been demonstrated. We observed that the J$_C$ increase with the increasing of the Al$_2$O$_3$ content in the matrix. We compared their superconducting and thermal properties with those obtained on a wire with the same architecture but with MONEL as external sheath and with a standard Cu Stabilized tape produced by Columbus Superconductors SpA. As we have seen the J$_C$ of AL60 is not too lower than MONEL one at 20 K both at low and high magnetic field, but unlike MONEL this new sample has a large amount of low resistivity material and a much higher thermal conductivity at all temperatures as well as the other two GlidCop® samples. In comparison with the standard Columbus tape, the AL60 showed a lower J$_C$ at magnetic fields up to 2T and a lower thermal conductivity at temperatures up to about 30K but higher at higher temperatures. This fact together with the good mechanical properties and a reasonable critical current density makes of GlidCop® sheathed wire a useful conductor in applications like Superconducting Fault Current Limiter which can operate at the liquid Neon temperature (about 27K) and where during the faults the temperature reaches values well above 30 K [16].